\documentclass[10pt,conference]{IEEEtran}
\IEEEoverridecommandlockouts
\usepackage{cite}
\usepackage{amsmath,amssymb,amsfonts}
\usepackage{graphicx}
\usepackage{textcomp}
\usepackage{cite}
\usepackage{url}
\usepackage[numbers,sort&compress]{natbib}
\usepackage{algorithm}
\usepackage{algpseudocode}
\usepackage{multirow}
\usepackage{amsmath}  
\usepackage{epstopdf}
\usepackage{comment}
\usepackage{booktabs}
\usepackage{makecell} 
\usepackage[table]{xcolor}
\usepackage{threeparttable}
\usepackage[a4paper, total={184mm,239mm}]{geometry}
\def\BibTeX{{\rm B\kern-.05em{\sc i\kern-.025em b}\kern-.08em
    T\kern-.1667em\lower.7ex\hbox{E}\kern-.125emX}}

\newcommand{\hongce}[1]{\textcolor{red}{[\textbf{hongce}: #1]}}

\begin{document}
\title{
\texttt{FORWORD}: Accelerating \underline{For}mal Datapath Verification via \underline{Word}-Level Sweeping
\thanks{This work is supported by National Natural Science Foundation of China, no. 62304194; and by the Guangdong S\&T Program no. 2025A0505000022.}
}


\author{
\IEEEauthorblockN{
Ziyi Yang\textsuperscript{1},
Guangyu Hu\textsuperscript{2},
Xiaofeng Zhou\textsuperscript{2},
Mingkai Miao\textsuperscript{1},
Changyuan Yu\textsuperscript{1},
Wei Zhang\textsuperscript{2},
Hongce Zhang\textsuperscript{1,2}
}
\IEEEauthorblockA{\textsuperscript{1} Hong Kong University of Science and Technology(Guangzhou)}
\IEEEauthorblockA{\textsuperscript{2} Hong Kong University of Science and Technology}
\IEEEauthorblockA{
\{zyang957, mmiao815, cyu418\}@connect.hkust-gz.edu.cn, 
\{ghuae, xzhoubu\}@connect.ust.hk, \{eeweiz, hongcezh\}@ust.hk
}
}

\maketitle

\begin{abstract}
Modern circuit design process increasingly adopts high-level hardware construction languages and parameterized design methodologies to shorten development cycles and maintain high reusability, in contrast to traditional hardware description languages. Such designs often involve complex datapath with arithmetic operations, wide bit-vectors, and on-chip memories, whose scale and level of modeling often pose significant challenges to formal datapath verification. Traditional bit-level SAT sweeping techniques lack the necessary abstraction and adaptability that are required to establish equivalence at a higher level.
%
In this paper, we propose \texttt{FORWORD}, a novel word-level sweeping verification engine tailored explicitly to formal datapath verification. \texttt{FORWORD} integrates randomized and constraint-driven word-level simulations, leveraging adaptive optimization to dynamically refine equivalent candidates identified during simulation. Experimental results demonstrate that \texttt{FORWORD} significantly outperforms state-of-the-art bit-level SAT sweeping engines and the monolithic SMT solving method, thanks to its enhanced capability in effectively identifying equivalent pairs. To the best of our knowledge, \texttt{FORWORD} is the first 
word-level sweeping engine explicitly designed for datapath verification, offering improved efficiency and adaptability to modern circuit designs.
\end{abstract}

\begin{IEEEkeywords}
formal methods, datapath verification, word-level sweeping, equivalence checking
\end{IEEEkeywords}

\section{Introduction}

Datapath verification is not merely a technical necessity but also a critical safeguard for modern digital circuit design, directly determining the correctness and reliability of hardware systems. In safety-critical domains, even subtle datapath flaws can cascade into system-wide failures, causing substantial damage. As circuits grow increasingly complex, driven by aggressive performance targets, sophisticated functionalities, and stringent power constraints, ensuring datapath correctness has become more and more challenging. 
The escalating complexity amplifies the risk of hidden logic errors, making efficient and scalable verification methodologies indispensable to ensure the integrity of contemporary hardware designs.

Modern hardware design frameworks such as Chisel~\cite{bachrach2012chisel}, SpinalHDL~\cite{Papon2017SpinalHDL} and high-level synthesis~\cite{coussy2010high} enable designers to build circuits from parameterized high-level modules that introduce intricate semantic constructs, which are inherently challenging for bit-level verification methods due to their limited abstraction capabilities and prohibitive computational overhead.\looseness=-1

SAT sweeping~\cite{mishchenko2005fraigs} has long served as an important verification method, which operates at the bit-level via And-Inverter Graphs (AIGs), taking advantage of structural hashing, simulation, and SAT solving.
It merges structurally identical subgraphs and use simulation to rapidly identify candidate equivalences, which are then refined by SAT checks, significantly reducing redundant logic and streamlines equivalence checking processes.
Despite its wide adoption, SAT sweeping faces fundamental limitations when verifying modern highly abstracted circuit designs, where semantics often extend beyond the bit level.

To address these challenges, verification has increasingly shifted from bit-level SAT reasoning to word-level Satisfiability Modulo Theories (SMT) solving. SMT solvers~\cite{niemetz2023bitwuzla,boolector,de2008z3,barbosa2022cvc5} inherently support reasoning at the word-level and handle complex semantics effectively. Nevertheless, existing SMT-based approaches lack the mechanisms of identifying equivalent nodes and the strategies of reducing the functional redundancy. To overcome this limitation, we propose a generalized word-level equivalence checking framework that supports a broader range of operators, combining the dynamic redundancy elimination strength of SAT sweeping with the abstract semantic reasoning capabilities of SMT solvers. This approach 
efficiently simplifies  circuit representations 
by recognizing word-level equivalences, thereby enabling more scalable datapath verification.

In summary, this paper makes the following contributions:
\begin{itemize}
\item We introduce \texttt{FORWORD}, a novel and scalable framework that implements a word-level sweeping method, providing efficient equivalence checking for modern datapath verification.
\item  We design a highly efficient method to generate simulation stimuli 
especially for the constrained circuits. Leveraging 
a combination of techniques including constraint analysis, simulated annealing and solution recombination, the method gives a good starting point to the random simulation in equivalence detection.
\item We develop a hybrid candidate reordering strategy within the \texttt{FORWORD} framework that significantly reduces SMT solver invocations, thereby improving the overall efficiency of the verification process.
\item We collect a comprehensive benchmark suite\footnote{https://github.com/yangziyiiii/FORWORD/tree/main/benchmark} for formal datapath verification, encompassing a diverse range of arithmetic designs to enable reproducible evaluation and foster further research in this domain.
\end{itemize}

This paper is structured as follows. 
Section~\ref{sec::background} presents the background of SAT sweeping and word-level circuit modeling. 
Section~\ref{sec::motivation} motivates word-level sweeping with a simple example.
Section~\ref{sec::method} introduces our new word-level equivalence checking method, with experimental evaluation in Section~\ref{sec::experiment}. 
Finally, we discuss related work in Section~\ref{sec::related} and conclude the paper in Section~\ref{sec::conclusion}.

\section{Background}
\label{sec::background}

\subsection{SAT Sweeping}
SAT sweeping has historically served as a foundational technique for combinational equivalence checking. It primarily operates on And-Inverter Graphs (AIGs) and employs techniques like structural hashing, bit-level simulation, and SAT solving. Structural hashing groups structurally identical subgraphs to simplify the circuit representations. Bit-level simulation assigns random input values to quickly identify potentially equivalent nodes based on the result of simulation, and SAT-based refinement employs SAT solvers to rigorously verify node equivalences. When the solver identifies a pair of nodes as inequivalent, the input assignment that can distinguish the pair is then integrated into subsequent simulations to refine node equivalence detection. Conversely, if the solver finds the constructed miter, a logical structure asserting the difference between two candidate nodes, to be unsatisfiable, it confirms their equivalence and the two subgraphs with distinct structures can then be unified. This helps to reduce circuit redundancy, compress the logic representation, and accelerate downstream formal verification tasks such as equivalence checking and property checking.

\subsection{BTOR2: a Word-Level Model}
\begin{figure}[t!]
\centering
\includegraphics[width=0.35\textwidth]{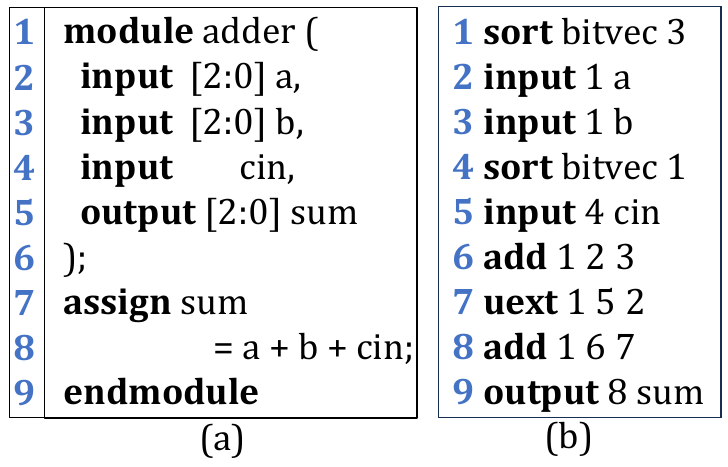}
\caption{Different circuit representations: (a) Verilog and (b) BTOR2}
\label{verilog2btor}
\end{figure}

The recent evolution of hardware modeling formats, exemplified by BTOR2~\cite{niemetz2018btor2}, underscores a shift toward word-level reasoning in design and verification. BTOR2 representation models circuits using the bit-vector and array theories. It supports rich constructs such as fixed-width logic and arithmetic operations, bit manipulations, and array read / write operations, making it well-suited for SMT-based verification. Tools like Yosys~\cite{wolf2013yosys} can be used to translate Verilog into BTOR2 models, preserving word-level semantics. This not only improves compatibility with SMT solvers such as Bitwuzla~\cite{niemetz2023bitwuzla}, but also enables higher-level reasoning during verification. An example of a circuit design represented using Verilog and BTOR2 format is shown by Figure~\ref{verilog2btor}.

\section{A Motivating Example}
\label{sec::motivation}

In datapath verification, traditional bit-level SAT sweeping has been widely used: signals are flattened into individual bits and equivalence is checked at that granularity. While effective for low-level redundancy removal, it discards high-level semantics such as arithmetic relations. The loss of such semantics often leads to harder SAT instances, increasing solver runtime and necessitating additional optimization strategies. To illustrate this limitation, 
let's consider the verification of a datapath with multiplication.

\begin{figure}[b]
\vspace{-1em}
\centering
\includegraphics[width=0.5\textwidth]{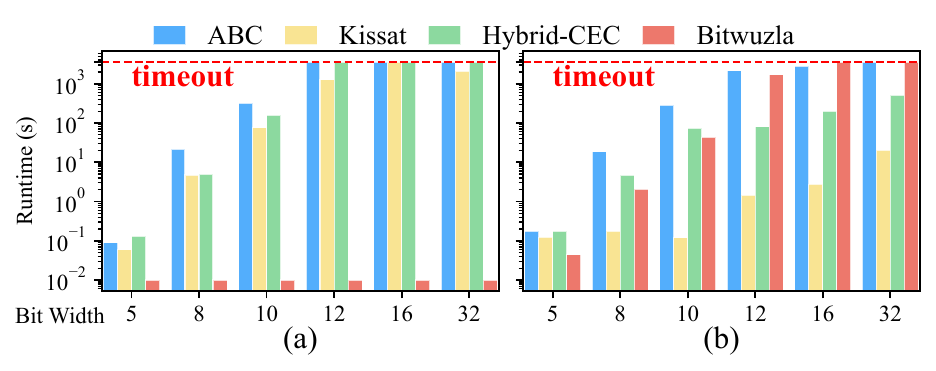}
\caption{Verification runtime in different scenarios when (a) word-level SMT solving prevails vs. (b) bit-level sweeping in Kissat is more effective} 
\label{pre-experiment}
\end{figure}

Generally speaking, multiplication is  usually difficult for SAT solvers, as shown by Figure~\ref{pre-experiment}(a), where we compare a range of state-of-the-art verification tools, including Bitwuzla~\cite{niemetz2023bitwuzla}, ABC $\&cec$ \cite{brayton2010abc}, Hybrid-CEC~\cite{HybridCEC} and Kissat~\cite{BiereFallerFazekasFleuryFroleyksPollitt-SAT-Competition-2024-solvers}. 
Except for Bitwuzla, all other tools work at the bit-level and are equipped with variants of SAT sweeping algorithms working either on the AIGs or on the CNF formulas. 
Bit-level tools such as \texttt{ABC} exhibit severe scalability degradation as bit-width increases. While verification of the 5-bit instance completes within milliseconds, the runtime for 8-bit and 10-bit instances escalates sharply, and for the 16-bit instance its runtime approaches the one-hour time limit.
On the other hand, \texttt{Bitwuzla} enjoys the benefit of a higher level of abstraction, whose runtime is not affected by the bit-width of the instances.

\begin{figure}[t]
\centering
\includegraphics[width=0.5\textwidth]{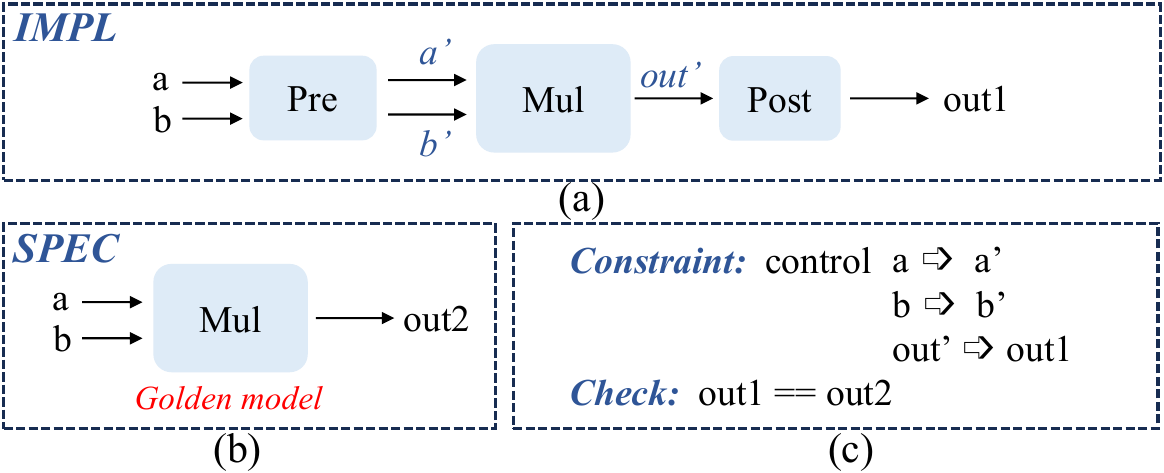}
\caption{An example of verifying a multi-functional unit under additional constraints}
\label{motivation}
\vspace{-1em}
\end{figure}

Despite the word-level abstraction, the state-of-the-art SMT solvers for bit-vectors, such as \texttt{Bitwuzla}, lack the mechanisms to identify and exploit equivalence information. In our applications, we often encounter multi-functional datapaths which can perform various related computations, for example, a multiplier with selective pre- and post-processing steps, such as signed or unsigned bit-width extension, masking, and result clamping, which are controllable by external inputs.
A simple example is shown in Figure~\ref{motivation}(a).
When verifying the functionality of such datapath, 
the verification engineers are usually provided with several golden models describing single computing functions in different use cases. Therefore, the equivalence checking tasks will take place under various additional constraints.
For example, Figure~\ref{motivation}(b) shows a single-functionality reference model which will be used to check the case when no additional pre- and post-processing are enabled. For a tool equipped with the sweeping technique (such as Kissat), it is able to identify internal equivalent points between the implementation and the specification under constraints and therefore, it takes the shortest time for this example. 

This example motivates the need for hybrid techniques that can combine the scalability of word-level reasoning with the strength of bit-level equivalence discovery, thereby addressing both high-level hardware modeling semantics and redundancy elimination in a unified framework.






\begin{figure*}[htbp]
\centering
\includegraphics[width=1\textwidth]{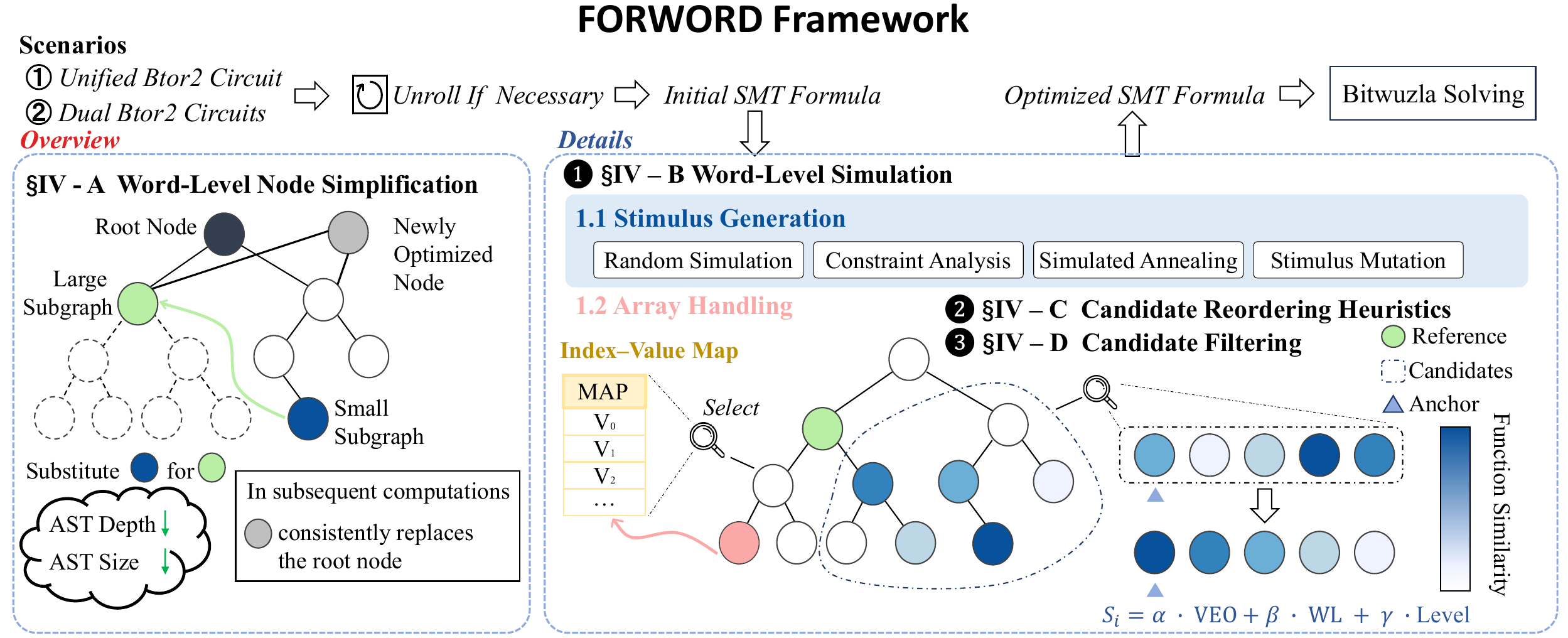}
\caption{Overview and details of the \texttt{FORWORD} flow.
The left side illustrates the overall simplification pipeline: larger subgraphs are progressively replaced with equivalent smaller ones, reducing AST depth and size, thereby shrinking the solver’s problem scope.
The right side details our simulation procedure: we use constraint- and heuristic-guided adaptive methods to generate stimuli, compute scores for candidate pairs, and apply a heuristic reordering policy followed by a lightweight filter so that the solver is invoked first on the most promising pairs. 
}
\label{framework}
\end{figure*}

\section{Methodology}
\label{sec::method}

This section presents our verification framework, \texttt{FORWORD}, which leverages word-level reasoning and various heuristics and strategies to address the scalability limitations of conventional bit-level verification approaches. Figure~\ref{framework} provides an overview of the framework.

\subsection{Overview of the FORWORD Framework}

Usually for datapath verification, an implementation circuit and a specification are provided to check for equivalence. They are either mitered in a unified BTOR2 (Scenario~1 in Figure~\ref{framework}) or are separated (Scenario~2).
For sequential circuits, unrolling may be necessary to obtain the complete input-output relation for verification. Then an initial SMT formula will be constructed, whose satisfiability indicates (in)equivalence. Additional constraints may also be provided that set the controlling input to check for conditional equivalence. Because typically the SMT formulas are structured as abstract syntax trees (ASTs) in the SMT solvers, in the following context, we will indistinguishably use SMT expressions and AST nodes to refer to the SMT formulas/terms.\looseness=-1

\texttt{FORWORD} performs word-level simplifications by unifying  equivalent sub-expressions. The overall algorithm is shown in Algorithm~\ref{alg:smt_sweep}.
It traverses the SMT formula representing the check in post order.
 During the traversal, \texttt{FORWORD} computes  simulation values and hash signatures for each node $t$ based on its children (Line 7-12).  This bottom-up traversal ensures that each node is evaluated only after all of its children are processed.
 For nodes with the same simulation results, \texttt{FORWORD} performs an SMT checking to validate their equivalence (Line 13-14). 
%
%
Confirmed equivalent nodes are merged and reused across the circuit (Line 15 and Line 6). Disproved pairs yield counter-examples, which are taken as new simulation vectors, refining quality of simulation over time (Line 18-20).
After completing the traversal, the framework applies the root-node substitution to simplify the original SMT formula before performing final check (Line 23-24). In the implementation, we don't explicitly compute structural hash to check for structural isomorphism. This is because the underlying SMT solver (i.e., Bitwuzla) is already equipped with such function that automatically unifies structurally isomorphic SMT terms.

In the following subsections,
We will highlight the features of \texttt{FORWORD} in more details.

\begin{algorithm}[ht]
\caption{The whole \texttt{FORWORD} Procedure}
\label{alg:smt_sweep}
\begin{algorithmic}[1]
\Require The SMT formula to check $r$, a constraint $\mathcal{C}_{\text{cons}}$
, initial simulation vector  $\mathcal{V}_{\text{sim}}$
\Ensure Verification result: $\texttt{SAT}$ or $\texttt{UNSAT}$

\State \textbf{Initialize:}
\State \hspace{1em} Simulation result $S : T \rightarrow \mathbb{B}^w$ ($T$ is the set of terms)
\State \hspace{1em} Hash table $\mathcal{H} : \mathbb{B}^w \rightarrow 2^T$
\State \hspace{1em} Substitution map $\theta : T \rightarrow T$
\ForAll{$t \in \text{PostOrder}(r)$}
    \State Substitute $t$'s children using $\theta$
    \If{$t$ is a leaf node}
        \State $S(t) \gets \mathcal{V}_{\text{sim}}(t)$
    \Else
        \State $S(t) \gets f_{\text{op}}(S(t_1), S(t_2), \dots)$ \Comment{Evaluate based on operator semantics and children $t_1,t_2,...$}
    \EndIf
    \State $\mathcal{H}(S(t)) \gets \mathcal{H}(S(t)) \cup \{t\}$

    \ForAll{$t' \in \mathcal{H}(S(t))$ where $t' \ne t$}
        \If{$\texttt{SMT}(\mathcal{C}_{\text{cons}} \wedge t \ne t') = \texttt{UNSAT}$}
            \State $\theta(t) \gets t'$ \Comment{Merge equivalent terms}
            \State \textbf{break}
        \Else 
            \State Extract model $\mu$ from solver
            \State $\mathcal{V}_{\text{sim}} \gets \mathcal{V}_{\text{sim}} \cup \mu$
            \State Update $S$ and $\mathcal{H}$
        \EndIf
    \EndFor
\EndFor
\State Apply substitution: $r \gets \theta(r)$
\State \Return $\texttt{SMT}(\mathcal{C}_{\text{cons}} \wedge r)$
\end{algorithmic}
\end{algorithm}





\subsection{Word-level Simulation}
The word-level simulation phase conducts constraint- and heuristic-guided adaptive simulation to efficiently identify potential equivalence relations among AST nodes.
While bit-level simulation on AIGs only needs to handle AND gates and inverters, word-level simulation in  \texttt{FORWORD} must support the full semantics of the bit-vector and array theory, including arithmetic operations like addition and subtraction, logic operations such as bit-wise AND and OR, bit manipulation such as concatenation and extraction, etc. \texttt{FORWORD} implements the simulation semantics of these operators following their formal definitions, ensuring accuracy across diverse circuit structures.


\subsubsection{Generation of Simulation Stimuli}
\texttt{FORWORD} begins with random simulation to identify sub-formulas that likely belong to the same equivalence class. However, for tightly constrained problems, purely random stimuli often fail to satisfy all constraints, leading to invalid input patterns that erroneously distinguish equivalent SMT expressions. 
To solve this problem, we propose the following strategies to generate constraint-aware simulation stimuli:
%
%

\vspace{0.5em}
\noindent
\textbf{Lightweight constraint analysis} captures simple constraint patterns such as $v_1=v_2$ (two input variables are equal) and $v_1=c_1$ (an input is set to some constants).  When generating input vectors for simulation, for the constrained variable pair in $v_1=v_2$, only one of $v_1$ or $v_2$ is given randomized values, while the other just takes its copy. For constraints in the form of $v_1=c_1$, $v_1$ will be forced to take the specified constant value. It enables targeted stimulus generation without significant overhead.

\vspace{0.5em}
\noindent
\textbf{Simulated annealing (SA) search and solution recombination} are used to handle complex constraint structures.
%
%
%
For complex constraints, we integrate a simulated annealing (SA) mutation process. Let $\mathcal{C}$ be the set of constraints and, for an input assignment $x\in\{0,1\}^n$, we define the satisfaction indicator $s_c(x)\in\{0,1\}$ for each $c\in\mathcal{C}$. The coverage of constraints and the corresponding energy function are defined as:\looseness=-1

\begin{equation}
\mathrm{cov}(x)=\tfrac{1}{|\mathcal{C}|}\sum_{c\in\mathcal{C}} s_c(x),
E(x)=|\mathcal{C}|-\sum_{c\in\mathcal{C}} s_c(x).
\end{equation}

We warm-start with an assignment $x^{(0)}$ that satisfies the simple equality/constant-binding constraints.
Then at iteration $k$, a neighbor $x'$ is generated from $x$ by position-biased bit-flip that toggles  a small number of bits drawn preferentially from variables appearing in currently unsatisfied constraints. Acceptance follows the Metropolis rule with a geometric cooling schedule:
\begin{equation}
\label{eq:sa-accept}
\begin{aligned}
& \Delta E := E(x') - E(x), \quad 0<\rho<1 \\
& P_{\mathrm{acc}} = \exp\!\big(-\max\{0,\Delta E\}/T\big), \quad T \leftarrow \rho\,T 
\end{aligned}
\end{equation}

The search terminates when a target coverage is reached or the budget on iterations is exhausted. 
To further diversify the input pattern, 
we also apply a lightweight mutation step: given two high-coverage solutions $x_a,x_b$ collected along the SA trajectory, we form $\tilde{x}=x_a \oplus x_b$ (bitwise XOR) as a new input pattern.
The constraint-aware stimuli generation consistently produces higher-quality input patterns than purely random simulation, with only a limited increase in runtime.

\subsubsection{Array Handling}
Arrays are commonly used in BTOR2 to model memory elements. In \texttt{FORWORD}, we build the index--value map to simulate arrays. 
The initial condition for an array is typically specified as layered write operations in BTOR2, which are extracted and recorded as the initial content of the index--value map. 

For array variables appear in the SMT formula, if any pair of them have the same index and element sort, we will compare their index--value maps. If the two contain identical entries for all indices, one will be substituted for the other via the global substitution map $\theta$. This substitution avoids maintaining redundant arrays with equivalent functional behavior, reducing representation sizes without requiring any SMT checking.

During simulation, we also make use of these stored index--value maps to evaluate the terms that reads from an array using the SMT ``\textit{select}" operator. For the SMT ``\textit{store}" operator, a temporary index--value map will be created to capture the write semantics.

\subsection{Candidate Reordering Heuristics}
To reduce the time required to validate pairs of equivalent candidates in Algorithm~\ref{alg:smt_sweep} (Lines 13–15), we reorder candidates that share identical simulation signatures using two lightweight structural similarities.

\noindent\textbf{Vertex-Edge Overlap (VEO): } 
As BTOR2 can be viewed as a directed acyclic graph (DAG) $G = (V,E)$, for the two SMT terms to compare, we first employ depth-first search (DFS) to systematically enumerate vertices and record edges of the two terms from their roots to leaves. To facilitate structural comparison, each vertex and edge is assigned a canonical hash signature. Since the exact computation of VEO can be costly, we approximate it via a weighted combination of vertex and edge Jaccard indices, yielding a scalable similarity measure.

Formally, let $V_1,V_2$ denote the vertex sets and $E_1,E_2$ denote the edge sets of two DAGs. The vertex overlap and edge overlap are defined using Jaccard indices, and the overall VEO score is obtained as their weighted combination:

\begin{equation}
\label{eq:veo-jaccard}
\begin{aligned}
&\mathrm{v\_overlap} =\frac{\lvert V_1 \cap V_2 \rvert}{\lvert V_1 \rvert + \lvert V_2 \rvert - \lvert V_1 \cap V_2 \rvert}\\
&\mathrm{e\_overlap} =\frac{\lvert E_1 \cap E_2 \rvert}{\lvert E_1 \rvert + \lvert E_2 \rvert - \lvert E_1 \cap E_2 \rvert}\\
&\mathrm{VEO} = \alpha\,\mathrm{v\_overlap} + (1-\alpha)\,\mathrm{e\_overlap}.
\end{aligned}
\end{equation}

\noindent \textbf{Position-aware WL features: }
Weisfeiler–Lehman (WL) refinement, a classical graph isomorphism heuristic, propagates labels across neighborhoods to summarize the \emph{structural context} of a root $r$ over increasing radii. This process yields a multi-scale fingerprint $\Phi_H(r)$ that encodes operator and bit-width patterns around $r$, independent of naming conventions. 

To compare two roots $r_1$ and $r_2$, we define the WL similarity as the weighted cosine between their fingerprints:
\begin{equation}
s_{\mathrm{WL}}(r_1,r_2) =
\frac{\big\langle \Phi_H(r_1),\,\Phi_H(r_2) \big\rangle}
     {\|\Phi_H(r_1)\|_2 \, \|\Phi_H(r_2)\|_2}.
\end{equation}

Intuitively, a high $s_{\mathrm{WL}}$ implies that the two cones exhibit similar local-to-midrange AST structure, making WL a permutation-invariant and name-agnostic indicator of subgraph resemblance. We use this metric to prioritize candidate pairs and prune the search space, while limiting its application on very large graphs where neighborhood aggregation becomes costly.\looseness=-1


\subsection{Equivalent Candidate Filtering}



When checking the candidates with the same simulation result, to avoid unnecessary solver overhead, we further apply a structural filter to quickly discard node pairs that are unlikely to match.
%
The filter checks the AST size of the two SMT terms. If the difference is greater than $5\times$, we will skip this pair and solver invocations are only triggered for candidates with high-confidence. Furthermore, when the terms in $\mathcal{H}(S(t))$ exceeds 100 candidates, rather than exhaustively validating all candidate pairs, the tool will sample a subset of candidates with the highest expected impact on structural reduction, measured by the difference in tree depth and the number of fan-outs. This approach ensures that solver resources are directed toward equivalences that offer the greatest simplification benefit.
A configurable threshold (i.e. 50 equivalent pairs per iteration) further bounds the number of solver invocations, maintaining scalability without compromising simplification quality. 

\begin{table*}[!t]
\caption{Results of experiments (runtimes are measured in seconds)
}
\centering
\scalebox{0.9}{
\begin{tabular}{ll c c c c c c c c c }
\toprule
\multicolumn{2}{l}{\textbf{Benchmark}} & \textbf{Max-Bitwidth} & \textbf{AST node size} & \textbf{PI/PO} & \texttt{ABC} & \texttt{Kissat} & \texttt{HybridCEC} & \texttt{Bitwuzla} & \textbf{Ours}  & \textbf{Speedup} \\
\midrule
\multirow{9}{*}{Uncond EC} 
 & AES\_comb           & 128 & 6396 & 494/1 & 23.99  & 249.40 &    $>3600$    & $>3600$ & 5.09  & 4.71\\
 & AES\_iter           & 128 & 12222 & 619/1 &$>3600$ & $>3600$ &   $>3600$   & $>3600$ & 3.12  & $>1153.84$\\
 & AES\_pipeline       & 128 & 11096 & 6281/1 & $>3600$ &  $>3600$ &  $>3600$ & $>3600$ & 3.10  & $>1161.29$ \\
 & HLS\_SEC\_1         & 32  & 166538 & 379/1 & 598.49  & $>3600$ &  $>3600$      & 68.97   & 20.45 & 3.37\\
 & HLS\_SEC\_2         & 32  & 216691 & 695/1 & 942.57  & $>3600$ &  $>3600$      & 6.83    & 2.33  & 2.93 \\
 & HLS\_SEC\_3         & 32  & 347748 & 462/1 & 837.79  & 2365.81 &  $>3600$      & 9.99    & 3.55  & 2.81\\
 & HLS\_SEC\_loop\_1   & 32  & 243820 & 436/1 & 626.27  & 3226.20 & $>3600$      & 7.79    & 2.35   & 3.31\\
 & HLS\_SEC\_loop\_2   & 32  & 179224 & 424/1 & 711.81  & $>3600$ & $>3600$       & 5.60    & 1.97  & 2.84\\
 & HLS\_SEC\_loop\_3   & 32  & 243387 & 632/1 & 1226.32 & $>3600$ & $>3600$       & 8.07    & 2.73  & 2.96\\
\midrule
\multirow{8}{*}{Cond-EC}
 & Mul\_modified       &  32  & 94 & 3/1 & $>3600$ &  52.87  & 593.11 & $>3600$ & 0.11 & 480.64\\
 & Div\_modified       &  32  & 94 & 3/1 & $>3600$ &  592.31 &  $>3600$ &  697.05 & 0.15 & 3948.73\\
 & SPI\_1              &  32 & 4895 & 650/1 &  82.90  &  16.46  &  58.01 &  20.50 &  13.17 & 1.25   \\
 & SPI\_2              &  32 & 4838 & 650/1 &  64.54  &  18.65  &  65.21 &  29.17  &  17.16 & 1.09   \\
 & SPI\_3              &  32 & 4853 & 650/1 &  71.43  &  22.08  &  54.28 &  17.78 & 13.35 & 1.33 \\
 & ILA\_Piccolo\_ADDI  &  32 & 75893 & 149/1 & $>3600$  & $>3600$  &  $>3600$ & 489.76  & 251.71 &  1.95  \\
 & ILA\_Piccolo\_SRA   &  32 & 84232 & 150/1 & $>3600$   & $>3600$ & $>3600$ &  671.62  & 299.70 &  2.24 \\
 & ILA\_Piccolo\_ORI   &  32 & 82946 & 149/1 &  $>3600$  & $>3600$  & $>3600$ & 1887.45 & 231.73 &  8.15 \\
 \midrule
    \multicolumn{3}{@{}l}{\textbf{GEOMEAN}} & & & & & & & &\textbf{11.11} \\
\bottomrule
\end{tabular}
}
\vspace{-2em}
\label{tab:benchmark}
\end{table*}

Leveraging the above techniques, \texttt{FORWORD} provides a word-level constraint-aware simulation framework for datapath verification. It systematically discovers, selectively checks and merges equivalent sub-terms to prune redundancy in the SMT formulation to reduce solver effort. The next section will report the resulting performance gains.

\section{Experiment}
\label{sec::experiment}

\subsection{Environment Setting}

We implement the proposed \texttt{FORWORD} framework in \texttt{C++} utilizing \texttt{SMT-Switch}~\cite{mann2021smt} to interface with the SMT solver \texttt{Bitwuzla}~\cite{niemetz2023bitwuzla}. All experiments are conducted on a server running Ubuntu 20.04.4 LTS, equipped with dual Intel Xeon Platinum 8375C processors and 256 GB of RAM. 

\subsection{Benchmark}
We evaluate the performance and scalability of \texttt{FORWORD} on a diverse set of hardware designs, comprising arithmetic datapath modules, cryptographic engines, designs generated from high-level synthesis, etc.
In total, the benchmark suite contains 511 cases. 
The evaluation targets two verification scenarios: (1) conditional equivalence checking (Cond-EC) and (2) unconditional equivalence checking (Uncond-EC).

\subsection{Tools in comparison}
For comparison, we include (1) the latest bit-level SAT sweeping method implemented in ABC~\cite{brayton2010abc}, invoked via the command ``\texttt{\&cec -m}'', (2) a hybrid combinational equivalence checking engine \texttt{Hybrid-CEC}~\cite{HybridCEC} that combines SAT sweeping and exact probability-based simulation, (3) the latest bit-level SAT solver \texttt{Kissat}~\cite{BiereFallerFazekasFleuryFroleyksPollitt-SAT-Competition-2024-solvers} and (4) monolithic SMT solving using the same  \texttt{Bitwuzla} solver. These tools represent state-of-the-art approaches at the bit and word levels.

\subsection{Results and Analysis}
 
\vspace{-1em}
\begin{figure}[htbp]
\centering
\includegraphics[width=0.4\textwidth]{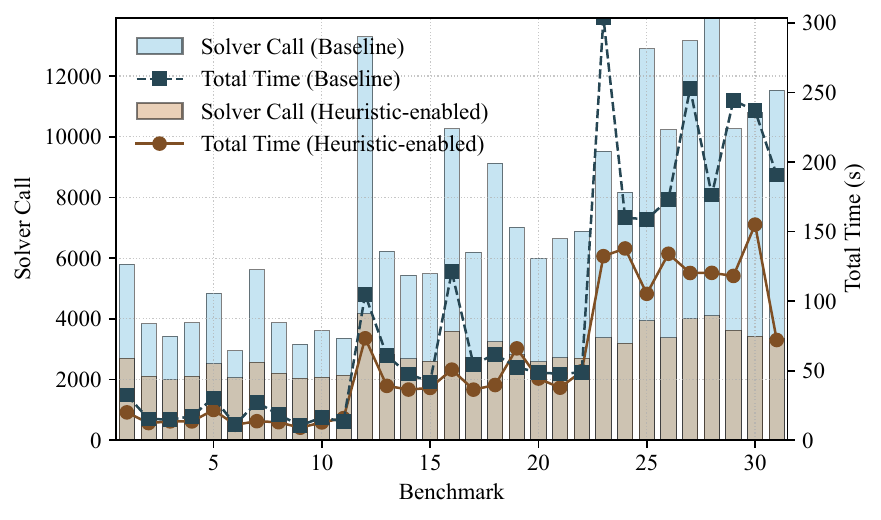}
\vspace{-1em}
\caption{Performance improvement using modern heuristic techniques}
\label{expr::heuristic}
\end{figure}

\begin{figure}[htbp]
\centering
\includegraphics[width=0.4\textwidth]{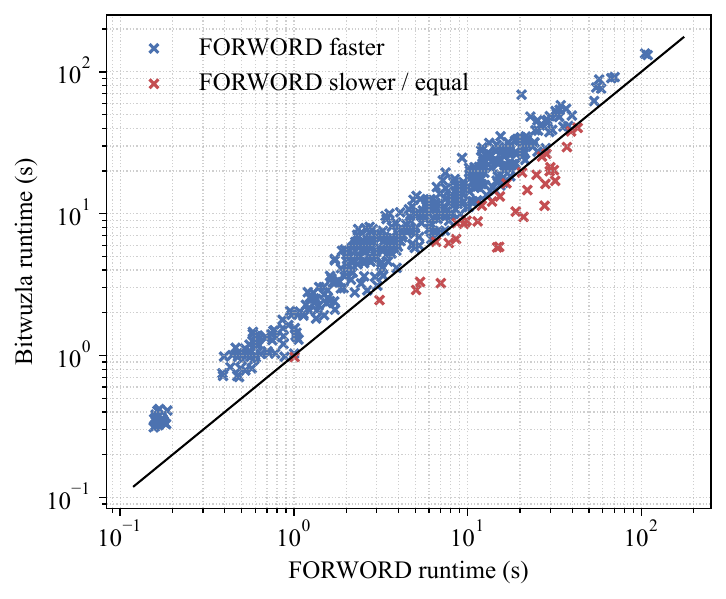}
\vspace{-1em}
\caption{Two word-level methods runtime comparison}
\vspace{-2em}
\label{expr::compare_bzla_sweeper}
\end{figure}

The experimental results are (partially) presented in Table~\ref{tab:benchmark}. 
Due to space limit,
we only report representative cases from each design set in the table.
The complete results are available at \url{https://figshare.com/s/e9ce05892872001f42f9}.
As circuits are mitered, there is only one primary output (PO), representing the (in)equivalence.
To provide an approximate characterization of circuit size and inherent solving difficulty, we list the maximum bit-width and the abstract syntax tree (AST) size.
The reported speedup denotes the ratio between the runtime of \texttt{FORWORD} and the fastest runtime among the four baseline tools.

We highlight several key observations:

\vspace{0.5em}
\noindent\textbf{Overall efficiency gains:} \texttt{FORWORD} achieves substantial runtime reductions across all benchmarks. On average, we observe an overall geometric mean speedup of 11.11$\times$ compared to the best-performing baseline. Notably, the runtime of \texttt{FORWORD} includes both the preprocessing stage and the final solving stage by \texttt{Bitwuzla}. In most cases, preprocessing is lightweight, typically completed within milliseconds and never exceeding two seconds.

\vspace{0.5em}
\noindent\textbf{Effectiveness of candidate reordering: } Our heuristic strategies significantly reduce the number of solver calls during the sweeping process. 
We conducted an ablation study with the heuristics disabled (the baseline) and enabled. 
As illustrated in Figure~\ref{expr::heuristic}, the heuristics lead to a reduction of approximately 60.62\% in total solver invocations and 63.06\% in cumulative solver runtime. This directly translates into notable improvements in end-to-end verification efficiency.

\vspace{0.5em}
\noindent\textbf{Advantages on datapath-intensive designs: } On large-scale and computation-intensive designs, such as the HLS\_SEC benchmarks, the benefits of word-level reasoning become particularly pronounced. These circuits contain a high density of arithmetic operators, which induce significant blow-ups in AIG and CNF encodings at the bit-level. By contrast, word-level reasoning preserves structural compactness, enabling both \texttt{Bitwuzla} and \texttt{FORWORD} to outperform bit-level tools by a wide margin. Moreover, \texttt{FORWORD} consistently identifies more equivalences than \texttt{Bitwuzla}, leading to further reductions in total solving time.

\vspace{0.5em}
\noindent\textbf{Scalability across benchmarks: } Figure~\ref{expr::compare_bzla_sweeper} compares solving times between the baseline \texttt{Bitwuzla} and our proposed \texttt{FORWORD} across all 511 test cases. The results show that \texttt{FORWORD} outperforms \texttt{Bitwuzla} on the majority of instances, confirming the scalability of our framework across a diverse range of datapath-oriented verification tasks.

\section{Related Work}
\label{sec::related}


At the bit level, recent efforts have also focused on strengthening simulation and exploiting parallelism. For example, SimGen~\cite{simgen} improves equivalence detection by generating simulation patterns via ATPG techniques rather than random vectors, reducing unnecessary SAT calls. ParallelCEC~\cite{possani2019parallel}
accelerates the classic CEC flow by partitioning miters and sweeping tasks across threads, achieving significant multi-core speedups compared to sequential baselines.

This paper focuses on bringing the sweeping technique into the word-level SMT solving, preserving arithmetic semantics that bit-level approaches lose. Unfortunately, ATPG-style pattern generation is not directly applicable in this setting, as there is no gate-level netlist at this higher abstraction level. Unlike ParallelCEC, which emphasizes parallel scalability, we focus on improving single-thread performance. These improvements are complementary---future parallel word-level sweeping engines can also benefit directly from stronger single-thread performance.\looseness=-1

\section{Conclusion}
\label{sec::conclusion}

This paper presents \texttt{FORWORD}, a word-level SMT simplification framework tailored for the verification of modern datapath designs. Experiments show that \texttt{FORWORD} effectively reduces solving time on a range of test cases, outperforming existing bit-level and word-level methods. 


\bibliographystyle{ieeetr}
\bibliography{IEEEabrv,ref}

\begin{thebibliography}{10}

\bibitem{bachrach2012chisel}
J.~Bachrach, H.~Vo, B.~Richards, Y.~Lee, A.~Waterman, R.~Avi{\v{z}}ienis, J.~Wawrzynek, and K.~Asanovi{\'c}, ``Chisel: constructing hardware in a scala embedded language,'' in {\em Proceedings of the 49th annual design automation conference}, pp.~1216--1225, 2012.

\bibitem{Papon2017SpinalHDL}
C.~Papon, ``{SpinalHDL: An alternative hardware description language}.'' Talk at FOSDEM 2017, Brussels, Belgium, Feb. 2017.

\bibitem{coussy2010high}
P.~Coussy and A.~Morawiec, {\em High-level synthesis}, vol.~1.
\newblock Springer, 2010.

\bibitem{mishchenko2005fraigs}
A.~Mishchenko, S.~Chatterjee, R.~Jiang, and R.~K. Brayton, ``{FRAIGs}: A unifying representation for logic synthesis and verification,'' tech. rep., ERL Technical Report, 2005.

\bibitem{niemetz2023bitwuzla}
A.~Niemetz and M.~Preiner, ``Bitwuzla,'' in {\em International Conference on Computer Aided Verification}, pp.~3--17, Springer, 2023.

\bibitem{boolector}
R.~Brummayer and A.~Biere, ``Boolector: An efficient smt solver for bit-vectors and arrays,'' in {\em Tools and Algorithms for the Construction and Analysis of Systems} (S.~Kowalewski and A.~Philippou, eds.), (Berlin, Heidelberg), pp.~174--177, Springer Berlin Heidelberg, 2009.

\bibitem{de2008z3}
L.~De~Moura and N.~Bj{\o}rner, ``Z3: An efficient smt solver,'' in {\em International conference on Tools and Algorithms for the Construction and Analysis of Systems}, pp.~337--340, Springer, 2008.

\bibitem{barbosa2022cvc5}
H.~Barbosa, C.~Barrett, M.~Brain, G.~Kremer, H.~Lachnitt, M.~Mann, A.~Mohamed, M.~Mohamed, A.~Niemetz, A.~N{\"o}tzli, {\em et~al.}, ``cvc5: A versatile and industrial-strength smt solver,'' in {\em International Conference on Tools and Algorithms for the Construction and Analysis of Systems}, pp.~415--442, Springer, 2022.

\bibitem{niemetz2018btor2}
A.~Niemetz, M.~Preiner, C.~Wolf, and A.~Biere, ``Btor2, {BtorMC} and {Boolector} 3.0,'' in {\em International Conference on Computer Aided Verification}, pp.~587--595, Springer, 2018.

\bibitem{wolf2013yosys}
C.~Wolf and J.~Glaser, ``Yosys-a free {Verilog} synthesis suite,'' in {\em Proceedings of the 21st Austrian Workshop on Microelectronics (Austrochip)}, vol.~97, 2013.

\bibitem{brayton2010abc}
R.~Brayton and A.~Mishchenko, ``{ABC}: An academic industrial-strength verification tool,'' in {\em Computer Aided Verification: 22nd International Conference, CAV 2010, Edinburgh, UK, July 15-19, 2010. Proceedings 22}, pp.~24--40, Springer, 2010.

\bibitem{HybridCEC}
Z.~Chen, X.~Zhang, Y.~Qian, Q.~Xu, and S.~Cai, ``Integrating exact simulation into sweeping for datapath combinational equivalence checking,'' in {\em 2023 IEEE/ACM International Conference on Computer Aided Design (ICCAD)}, pp.~1--9, 2023.

\bibitem{BiereFallerFazekasFleuryFroleyksPollitt-SAT-Competition-2024-solvers}
A.~Biere, T.~Faller, K.~Fazekas, M.~Fleury, N.~Froleyks, and F.~Pollitt, ``{CaDiCaL}, {Gimsatul}, {IsaSAT} and {Kissat} entering the {SAT Competition 2024},'' in {\em Proc.~of {SAT Competition} 2024 -- Solver, Benchmark and Proof Checker Descriptions} (M.~Heule, M.~Iser, M.~J{\"a}rvisalo, and M.~Suda, eds.), vol.~B-2024-1 of {\em Department of Computer Science Report Series B}, pp.~8--10, University of Helsinki, 2024.

\bibitem{mann2021smt}
M.~Mann, A.~Wilson, Y.~Zohar, L.~Stuntz, A.~Irfan, K.~Brown, C.~Donovick, A.~Guman, C.~Tinelli, and C.~Barrett, ``{SMT}-switch: a solver-agnostic {C++} {API} for {SMT} solving,'' in {\em International Conference on Theory and Applications of Satisfiability Testing}, pp.~377--386, Springer, 2021.

\bibitem{simgen}
C.~Rizzi, S.~Brunner, A.~Mishchenko, and L.~Josipović, ``Simgen: Simulation pattern generation for efficient equivalence checking,'' in {\em 2025 Design, Automation \& Test in Europe Conference (DATE)}, pp.~1--7, 2025.

\bibitem{possani2019parallel}
V.~N. Possani, A.~Mishchenko, R.~P. Ribas, and A.~I. Reis, ``Parallel combinational equivalence checking,'' {\em IEEE Transactions on Computer-Aided Design of Integrated Circuits and Systems}, vol.~39, no.~10, pp.~3081--3092, 2019.

\end{thebibliography}
\end{document}